# A Search for Interstellar Carbon Chain Alcohol $HC_4OH$ in Star-Forming Region L1527 and Dark Cloud TMC-1

Short Title: A Search for $HC_4OH$ in L1527 and TMC-1

MITSUNORI ARAKI,[1] SHURO TAKANO,[2,3] HIROMICHI YAMABE,[1] NAOHIRO KOSHIKAWA,[1] KOICHI TSUKIYAMA,[1] AYA NAKANE,[4] TOSHIAKI OKABAYASHI,[4] ARISA KUNIMATSU[5] AND NOBUHIKO KUZE[5]

[1]Department of Chemistry, Faculty of Science Division I, Tokyo University of Science, 1-3 Kagurazaka, Shinjuku-ku 162-8601, Tokyo, Japan; araki@rs.kagu.tus.ac.jp

[2]Nobeyama Radio Observatory, 462-2 Nobeyama, Minamimaki, Minamisaku, Nagano 384-1305, Japan

[3]Visiting researcher: Leiden Observatory, Leiden University, P.O. Box 9513, NL-2300 RA, Leiden, The Netherlands

[4]Department of Chemistry, Faculty of Science, Shizuoka University, 836 Oya, Suruga-ku, Shizuoka 422-8529, Japan

[5]Department of Materials and Life Sciences, Faculty of Science and Technology, Sophia University, 7-1 Kioi-cho, Chiyoda-ku 102-8554, Tokyo, Japan

**ABSTRACT**

We report a sensitive search for the rotational transitions of the carbon chain alcohol $HC_4OH$ in the frequency range of 21.2–46.7 GHz in the star-forming region L1527 and the dark cloud TMC-1. The motivation was laboratory detection of $HC_4OH$ by microwave spectroscopy. Despite achieving rms noise levels of several millikelvin in the antenna temperature using the 45 m telescope at Nobeyama Radio Observatory, the detection was not successful, leading to 3σ upper limits corresponding to the column densities of $2.0 \times 10^{12}$ and $5.6 \times 10^{12}$ $cm^{-2}$ in L1527 and TMC-1, respectively. These upper limits indicate that $[HC_4OH]/[HC_5N]$ ratios are less than 0.3 and 0.1 in L1527 and TMC-1, respectively, where $HC_5N$ is an $HC_4$-chain cyanide and $HC_4OH$ is a hydroxide. These ratios suggest that the cyano carbon chain molecule dominates the hydroxyl carbon chain molecule in L1527 and TMC-1. This is contrary to the case of saturated compounds in hot cores, *e.g.*, $CH_3OH$ and $CH_3CN$, and can be a chemical feature of carbon chain molecules in L1527 and TMC-1. In addition, the column densities of the "unsubstituted" carbon chain molecule $C_4H$ and the sulfur-bearing molecules SO and $HCS^+$ were determined from detected lines in L1527.

## 1. INTRODUCTION

The abundance ratio between two related molecules can contribute to the study of an interstellar cloud by serving as an effective probe of a chemical feature and a chemical reaction. The ratio between hydroxyl and cyano compounds with the same molecular frame, [*X*OH]/[*X*CN], is one of the most interesting parameters. For saturated compounds, the ratios in hot cores and dark clouds have been reported. For hot cores, the [$CH_3OH$]/[$CH_3CN$] ratios were observed to be 70 and 280, and the [$CH_3CH_2OH$]/[$CH_3CH_2CN$] ratios were 23 and 13 in Sgr B2 and Orion-KL, respectively, *i.e.,* the hydroxyl compounds are abundant compared to the cyano compounds (Turner 1991). For the dark cloud, the [$CH_3OH$]/[$CH_3CN$] ratio in TMC-1 was observed to be 4 (Irvine et al. 1987). Although the ratios depend on the cloud and cores, the hydroxyl compound is generally dominant, which can be one of the chemical features of saturated compounds.

Because linear carbon chain molecules are abundant in some dark clouds and low-mass star-forming regions (*e.g.*, TMC-1 and L1527, respectively), the [*X*OH]/[*X*CN] ratio for linear carbon chain molecules is interesting. In these cases, *X*CN is typically a cyanopolyyne ($HC_{2n}CN$: $n$ = 1, 2, 3 …); therefore, [*X*OH]/[*X*CN] can be expressed as [$HC_{2n}OH$]/[$HC_{2n}CN$]. For the shortest carbon chain, *i.e.*, $n$ = 1, $HC_{2n}OH$ corresponds to $HC_2OH$; however, its spectral lines have not yet been measured in the laboratory. Therefore, we are interested in the ratio for $n$ = 2 (*i.e.*, [$HC_4OH$]/[$HC_4CN$]), where $HC_4OH$ is the carbon chain alcohol, H–C≡C–C≡C–OH. Hereafter, $HC_4CN$ is written as $HC_5N$.

The first laboratory observation of $HC_4OH$ was carried out in the 17–39 GHz region by the Stark modulation microwave spectrometer at Sophia University (Araki and Kuze 2008). The $K_a$ = 0 transitions were not observed in this observation and the $K_a$ = 1 transitions were misassigned as the $K_a$ = 0 transitions. The correct $K_a$ = 0 transitions were detected with the Fourier transform microwave spectrometer at Shizuoka University (Kuze et al. in preparation). The precise rest frequencies of the $K_a$ = 0 and $K_a$ = 1 transitions were measured in the 12–47 GHz region.

To detect $HC_4OH$, a target cloud should be rich in carbon chain and oxygen-bearing molecules. Many carbon chain molecules were detected in the star-forming region L1527 (*e.g.*, Sakai et al. 2008a) and the dark cloud TMC-1 (*e.g.*, Kaifu et al. 2004). Oxygen-bearing molecules such as HCNO (Marcelino et al. 2009), HNCO (Marcelino et al. 2009), $HOCO^+$ (Sakai *et al*. 2008b and 2008c), $H_2CO$ (Dickens and Irvine 1999), $CH_3OH$ (Sakai *et al*. 2009), $HCO^+$ (Gregersen et al. 1997), and CO (*e.g.*, Fuller et al. 1996) were detected in L1527. Thus, the detection of these molecules suggests that L1527 may be one of the best targets to detect the carbon chain alcohol $HC_4OH$. Therefore, we have selected L1527 as the primary detection source for $HC_4OH$ and TMC-1 as the secondary detection source. Irrespective of the detection of $HC_4OH$, the present work allows us to establish a chemical feature of carbon chain molecules in L1527 and TMC-1.

Using the precise rest frequencies (Kuze et al.), we have searched for $HC_4OH$ in L1527 and TMC-1 with the 45 m telescope at Nobeyama Radio Observatory (NRO). Carbon chain molecules including $HC_5N$, $C_4H$, and $HC_7N$ and sulfur-bearing molecules including SO and $HCS^+$ were detected in the present observation. In this paper, we report the upper limits of the column density of $HC_4OH$ and compare its abundance with related species.

## 2. OBSERVATIONS

Observations of the molecular lines in the 21, 29–47, and 80–90 GHz regions were carried out with the NRO 45 m telescope[1] on January 15–21, 2010, and March 4 and 5, 2011, for L1527 and April 11–17, 2009, for TMC-1. We observed the IRAS 04368+2557 position in L1527 (α2000.0, δ2000.0) = ($04^h$ $39^m$ $53^s.89$, 26°03′11.0″) and the cyanopolyyne peak in TMC-1 (α2000.0, δ2000.0) = ($04^h$ $41^m$ $42^s.49$, 25°41′27.0″). The HEMT receivers H22 and H30

[1] The 45m telescope is operated by Nobeyama Radio Observatory, a branch of National Astronomical Observatory of Japan.

and the SIS mixer receivers S40 and T100V were used for the observations; their typical single sideband system temperatures were about 150, 500, 220, and 200 K for H22, H30, S40 and T100V, respectively. The main beam efficiencies were 0.83, 0.79, 0.73, 0.72, and 0.42 in the 21, 29–35, 35–40, 40–47, and 80–90 GHz regions, respectively, which were not further corrected for slight changes in frequencies. The beam sizes of the telescope are 79″, 55″, 39″, and 19″ at 21, 30, 43, and 80 GHz, respectively. The telescope pointing was verified by observing the nearby SiO maser source NML Tau every 90 min. The typical pointing accuracy was a few arcseconds. The position switching mode was employed for the observations, where the off position was taken at $\Delta\alpha$ = 30′ and $\Delta\delta$ = 30′. A set of acousto-optical radio spectrometers (AOSs) was used for the backend. Eight wide band AOSs have individual bandwidths of 250 MHz, and 8 high resolution AOSs have individual bandwidths of 40 MHz. The velocity resolution of the former is 1.87 km $s^{-1}$ at 40 GHz whereas that of the latter is 0.28 km $s^{-1}$. The calibration of the intensity scale was carried out by the chopper wheel method.

## 3. RESULTS AND DISCUSSION

### 3.1. *Carbon Chain Alcohol $HC_4OH$*

To estimate the distribution of $HC_4OH$ in L1527, we measured the intensity of the transition of $HC_5N$, which has a comparable carbon chain size. The $J$ = 16–15 transition in the 42.6 GHz region observed using the Green Bank Telescope (GBT) with a beam size of 17.5″ was reported by Sakai et al. (2009). In the present observation, the $J$ = 16–15 transition was observed again using the NRO telescope, as shown in the spectrum (a) of Figure 1. The observed integrated intensity (denoted as $W$ in Table 1) of the transition using the NRO telescope was 72% of that using the GBT. For the NRO telescope, the lower intensity can be due to beam dilution. The distribution of $HC_5N$ is estimated to be 33″, which is less than the beam sizes for the present $HC_4OH$ search. To determine the column density of $HC_4OH$ in L1527, the source coupling factor was assumed on the basis of the distribution of $HC_5N$. By comparison, the distribution of $HC_5N$ is smaller than that of the "unsubstituted" carbon chain molecule $C_4H$, because the distribution of $C_4H$ was reported to be extended over a 40″ scale around the center of L1527 (Sakai et al. 2008a).

First, we searched for the $J_{Ka,Kc} = 7_{0,7}–6_{0,6}$, $9_{0,9}–8_{0,8}$, $10_{0,10}–9_{0,9}$, and $11_{0,11}–10_{0,10}$ transitions (29.7–46.7 GHz region) of $HC_4OH$ in L1527 and the $J_{Ka,Kc} = 5_{0,5}–4_{0,4}$ and $7_{0,7}–6_{0,6}$ transitions (21.2 and 29.4 GHz, respectively) of $HC_4OH$ in TMC-1 by using the wide band AOSs. No lines were detected in L1527 and TMC-1, as shown in Table 1, despite achieving rms noise levels of several millikelvin for antenna temperature $T_A^*$. The upper limits of column densities of $HC_4OH$ in L1527 and TMC-1 were determined to be $4.1 \times 10^{12}$ and $5.6 \times 10^{12}$ $cm^{-2}$, respectively, on the basis of the assumption of the excitation temperatures of 12.3 and 3.8 K, respectively ($C_4H_2$, Sakai et al. 2008a), using 3σ rms noise and the LTE assumption. Partition functions were calculated by summing over the energy levels of the $J$ = 0 to 31 states for $K_a$ = 0 and $K_a$ = 1, where the energy levels for $K_a$ = 2 have negligible contributions to the partition functions at the temperatures of L1527 and TMC-1. The calculated dipole moment $\mu_a$ = 1.65 D on the a-axis at the B3LYP/cc-pVQZ level reported by Araki & Kuze (2008) was used.

Second, we intensively searched for the $J_{Ka,Kc} = 10_{0,10}–9_{0,9}$ transition of $HC_4OH$ in L1527 using the high resolution AOS, as shown in the spectrum (b) of Figure 1, because this transition is the most intense at the excitation temperature of 12.3 K. The rms noise level reached 9 mK at $T_A^*$. The spectrum (b) gave the stricter upper limit for the column density of $2.0 \times 10^{12}$ $cm^{-2}$ in L1527.

For $HC_5N$, Sakai et al. (2008a, 2009) reported not only the $J$ = 16–15 transitions but also the $J$ = 7–6, 17–16, and 32–31 transitions in L1527, and the column density and the excitation temperature were reported to be $(6.8 \pm 1.4) \times 10^{12}$ $cm^{-2}$ and 14.7 ± 5.3 K, respectively. Takano et al. (1990) reported the column density of $HC_5N$ to be $(6.3 \pm 0.6) \times 10^{13}$ $cm^{-2}$ in TMC-1. The upper limits of $HC_4OH$ obtained in this work indicate that the $[HC_4OH]/[HC_5N]$ ratios are less than 0.3 and 0.1 in L1527 and TMC-1, respectively.

The $[HC_4OH]/[HC_5N]$ ratios are results of production and destruction reactions. The OH and CN molecules are possible keys for the formation of $HC_4OH$ and $HC_5N$, respectively. Dipole moments of OH and CN are one of effective parameters to determine rate constants of ion-molecular reactions, which may be a part of their production reactions. However the dipole moments cannot give a simple explanation about the $[HC_4OH]/[HC_5N]$ ratios, because the dipole moment of CN is comparable with that of OH. In addition, the reactions to produce $HC_4OH$ and $HC_5N$ may be governed by a difference in the abundance between CN and OH. However the column densities of OH [$2.6 \times 10^{15}$ $cm^{-2}$ (Harju et al. 2000)] in TMC-1 is larger than that of CN [$2.9 \times 10^{14}$ $cm^{-2}$ (Crutcher et al. 1984)]. Thus the column densities of CN and OH cannot give a simple explanation as well as the dipole moments.

These $[HC_4OH]/[HC_5N]$ ratios in the present observation suggest that the cyano carbon chain molecule dominates the hydroxyl carbon chain molecule in L1527 and TMC-1, which is contrary to the case of saturated compounds in hot cores, *i.e.*, the $[CH_3OH]/[CH_3CN]$ ratios in Sgr B2 and Orion-KL were 70 and 280, respectively, and the $[CH_3CH_2OH]/[CH_3CH_2CN]$ ratios are 23 and 13, respectively (Turner, 1991). $CH_3CN$ can be produced by gas phase reactions affected by grain mantle evaporation (Kaleuskii et al. 2000). Some chemical reaction models for the saturated compounds demonstrated that $CH_3CH_2CN$ (Miao & Snyder 1997), $CH_3OH$ and $CH_3CH_2OH$ (Charnley et al. 1992) can be produced by grain surface reactions. Therefore, the $[HC_4OH]/[HC_5N]$ ratios may contribute in solving reaction chemistry of $HC_4OH$ .

### 3.2 "*Unsubstituted*" *Carbon Chain Molecule* $C_4H$

In L1527, the $N_J = 4_{4.5}$–$3_{3.5}$ and $4_{3.5}$–$3_{2.5}$ transitions of the "unsubstituted" carbon chain molecule $C_4H$ were observed in the 38.1 GHz region, as shown in the spectrum (e) of Figure 1, and the $9_{9.5}$–$8_{8.5}$ and $9_{8.5}$–$8_{7.5}$ (85.6 and 85.7 GHz, respectively) transitions reported by Sakai et al. (2008a) were observed again. The column density and the excitation temperature of $C_4H$ were simultaneously determined from the four lines to be (1.01 ± 0.05) × $10^{14}$ $cm^{-2}$ and 14.3 ± 1.3 K, respectively, by using a dipole moment of 0.87 D (Woon 1995), in which the errors are one standard deviation. This excitation temperature is comparable with that of $C_4H_2$ (12.3 ± 0.8 K) (Sakai et al. 2008a). The column density of $C_4H$ in L1527 is one third of 2.9 × $10^{14}$ $cm^{-2}$ reported by Sakai et al. (2008a) in TMC-1. The [$C_4H$]/[$HC_5N$] ratios of 4.6 in TMC-1 and 14.8 in L1527 indicate that $C_4H$ is considerably more abundant than $HC_5N$ as well as $HC_4OH$. Recently, Sakai and coworkers (2008a) concluded that $C_nH_m$ molecules in L1527 are exceptionally abundant among the various series of carbon chain molecules, and that carbon chain cyanide molecules H–$(CC)_n$–CN ($n$ = 2, 3, 4) in L1527 are less abundant compared with those in TMC-1. The [$C_4H$]/[$HC_5N$] ratio in L1527 obtained in this work agrees with their conclusion.

For the chemical reactions to produce $HC_5N$, the ion-molecular reactions *via* $C_4H$ and the neutral-neutral reactions, $C_4H_2$ + CN, were suggested by Millar & Freeman (1984) and Fukuzawa et al. (1998), respectively. The [$C_4H$]/[$HC_5N$] ratios contribute to the knowledge of the chemical reactions that produce $HC_5N$. The [$C_4H$]/[$HC_4OH$] ratios, which are larger than 52 in TMC-1 and L1527, constitute useful information for understanding the production of $HC_4OH$.

### 3.3. *Sulfur-bearing Molecules SO and $HCS^+$*

The $J$ = 2–1 and 1–0 transitions of $HCS^+$ in TMC-1 have been reported (Thaddeus *et al.*, 1981; Irvine et al. 1983; Hirahara et al. 1992). The resultant column density of $HCS^+$ in TMC-1 was determined to be (1.0 ± 0.3) × $10^{13}$ $cm^{-2}$ (Hirahara et al. 1992). In the present work, $HCS^+$ was weakly detected in L1527 for the first time, as shown in the spectrum (d) of Figure 1. By using the detected $J$ = 1–0 transition, the column density of $HCS^+$ in L1527 was determined to be (5.9 ± 1.8)×$10^{11}$ $cm^{-2}$ on the basis of the fixed excitation temperature of 12.3 K and the dipole moment of 1.958 D, which was obtained by an *ab initio* calculation (Botschwina and Sebald 1985). For the column density determined by a single line, the resultant error is evaluated from uncertainties in the integrated intensity and the excitation temperature (±2.3 K, the 3σ error for $C_4H_2$) (Sakai et al. 2008a). The column density of $HCS^+$ in L1527 is significantly less than that in TMC-1 [(1.0 ± 0.3) × $10^{13}$ $cm^{-2}$] (Hirahara et al. 1992). The [$HCS^+$]/[CS] ratio is an index parameter for the dissociative recombination of sulfur bearing molecules in the chemical model calculation (Millar & Herbst 1990, Montaigne et al. 2005). For the dark cloud, the observed ratio of 0.044 at the cyanopolyyne peak in TMC-1 (Hirahara et al. 1992) was larger than the calculated ratio of 0.015 by Millar and Herbst (1990) and those of 0.01-0.001 by Montaigne et al. (2005). For the low-mass star-forming region, the observed ratio in L1527 was determined to be 0.06 using the column density of $HCS^+$ in the present observation and that of CS reported by Jørgensen et al. (2002, 2004). Thus the observed ratio is considerably larger than the calculated ratios for the dark clouds. L1527 and TMC-1 have the comparable ratios in despite of different excitation temperatures.

The $N_J = 0_1–1_0$ transition of SO was already detected in TMC-1 and the column density was determined to be $5×10^{13}$ $cm^{-2}$ (Rydbeck et al. 1980). In the present work, the same SO transition was detected in L1527, as shown by the spectrum (c) of Figure 1. The column density in L1527 was found to be (5.3 ± 1.1)×$10^{13}$ $cm^{-2}$ for the fixed excitation temperature of 12.3 K and the dipole moment of 1.52 D (Lovas et al. 1992). The column density in L1527 is comparable with that in TMC-1. The observed SO line has possible wing features due to the outflow at 5–7.5 km $s^{-1}$ as well as the reported $J$ = 1–0 line of $HCO^+$ (Sakai et al. 2008a).

Hirota et al. (2001) found out that the abundance of CCS in L1527 is less than that in TMC-1. Sakai et al. (2008a) reported that the abundances of CCS, $HC_5N$, $HC_7N$ and $HC_9N$ are lower in L1527 than those in TMC-1, and that the $C_nH_m$ (n ≥ 3, m ≥ 1) species seem to be exceptionally abundant among various series of the carbon-chain molecules. Sakai et al. (2008a) mentioned the possibility of a long life time and regeneration of the $C_nH_m$ species. In the present work, similarly, the triatomic molecule $HCS^+$ in L1527 is less abundant than that in TMC-1 although the diatomic molecule SO has comparable abundance. Low abundances in L1527 are not only for the carbon chain alcohol $HC_4OH$ (upper limits) and the carbon chain cyanide $HC_5N$ but also of the triatomic sulfur bearing molecules $HCS^+$ and CCS.

The authors thank Prof. Satoshi Yamamoto and Dr. Nami Sakai at the University of Tokyo for their helpful comments on the assignment of $HC_4OH$ and the analysis of the column density. S.T. thanks Prof. E.F. van Dishoeck and Leiden Observatory for hosting during his stay. S.T. also thanks Yamada Science Foundation and Netherlands Research School for Astronomy (NOVA) for financial support.

**FIGURE LEGEND**

Figure 1. Spectral line profiles observed in L1527.

These spectra were obtained under the assumption that $V_{LSR}$ for L1527 is 5.83 km $s^{-1}$ ($C_6H$) (Sakai et al. 2007). The spectra (a), (b), (c) and (d) were acquired with high resolution AOSs and the spectrum (e) was acquired with a wide band AOS.

Table 1. Molecular lines Observed in L1527 and TMC-1.

| Species | Transition | Frequency [a] (GHz) | $T_A^*$ (K) | $dv$ [b] (km $s^{-1}$) | $W$ [c] (K km $s^{-1}$) | rms (mK) | $V_{LSR}$ (km $s^{-1}$) | Ref.[d] |
|---|---|---|---|---|---|---|---|---|
| | | | | L1527 | | | | |
| $HC_4OH$ | $J_{Ka,Kc} = 7_{0,7}–6_{0,6}$ | 29.749313[e] | <0.024[f] | 2.5[g,h] | <0.079 | 7.9 | - | |
| | $J_{Ka,Kc} = 9_{0,9}–8_{0,8}$ | 38.245234[e] | <0.019[f] | 2.2[g,h] | <0.063 | 6.4 | - | |
| | $J_{Ka,Kc} = 10_{0,10}–9_{0,9}$ | 42.492143[e] | <0.027[f] | 0.60[h] | <0.024 | 9.0 | - | |
| | $J_{Ka,Kc} = 11_{0,11}–10_{0,10}$ | 46.738243[e] | <0.037[f] | 2.0[g,h] | <0.110 | 12.3 | - | |
| $C_4H$ | $N_J = 4_{4.5}–3_{3.5}$ | 38.049655[i,j] | 0.203(6) | 2.36(6)[g] | 0.70(4) | 5.3 | 6.4 | |
| | $N_J = 4_{3.5}–3_{2.5}$ | 38.088461[i,j] | 0.165(10) | 1.89(13)[g] | 0.45(6) | 5.3 | 6.3 | |
| | $N_J = 9_{9.5}–8_{8.5}$ | 85.634010[i,j] | 0.371(7) | 1.12(2)[g] | 1.05(4) | 4.2 | 5.8 | 1 |
| | $N_J = 9_{8.5}–8_{7.5}$ | 85.672580[i,j] | 0.336(7) | 1.12(3)[g] | 0.95(4) | 4.2 | 5.8 | 1 |
| $HC_5N$ | $J = 10–9$ | 29.289152[k] | 0.332(10) | 0.579(19) | 0.259(16) | 9.3 | 5.88 | |
| | $J = 16–15$ | 42.602154[l] | 0.558(6) | 0.518(6) | 0.422(9) | 6.4 | 5.87 | 2 |
| $HC_7N$ | $J = 34–33$ | 38.351449[m] | 0.029(5) | 1.6(5)[g] | 0.07(3) | 12.3 | 6.0 | |
| SO | $N_J = 0_1–1_0$ | 30.001543[n] | 0.613(17) | 0.81(3) | 0.67(4) | 13.9 | 6.02 | |
| $HCS^+$ | $J = 1–0$ | 42.674195[j] | 0.052(7) | 0.43(7) | 0.033(10) | 8.4 | 5.88 | |
| | | | | TMC-1 | | | | |
| $HC_4OH$ | $J_{Ka,Kc} = 5_{0,5}–4_{0,4}$ | 21.251128[e] | <0.009[f] | 2.7[g,h] | <0.032 | 2.9 | - | |
| | $J_{Ka,Kc} = 7_{0,7}–6_{0,6}$ | 29.749296[e] | <0.031[f] | 2.5[g,h] | <0.103 | 10.3 | - | |

[a] Rest frequency.

[b] Obtained by Gaussian fit.

[c] $W = \int \frac{T_A^*}{\eta_{mb}} dv$, where $\eta_{mb}$ is the main beam efficiency.

[d] References are for the previous observations ([1]Sakai et al. 2008a; [2]Sakai et al. 2009).

[e] Rest frequencies were derived from Kuze et al.

[f] 3σ upper limit.

[g] Wide band AOS.

[h] To determine $W$, typical line widths in frequencies were assumed.

[i] Averages of two hyperfine components.

[j] The Cologne Database for Molecular Spectroscopy.

[k] Alexander et al. 1976.

[l] Winnewisser et al. 1982.

[m] Kirby et al. 1980.

[n] Kaifu et al. 2004.

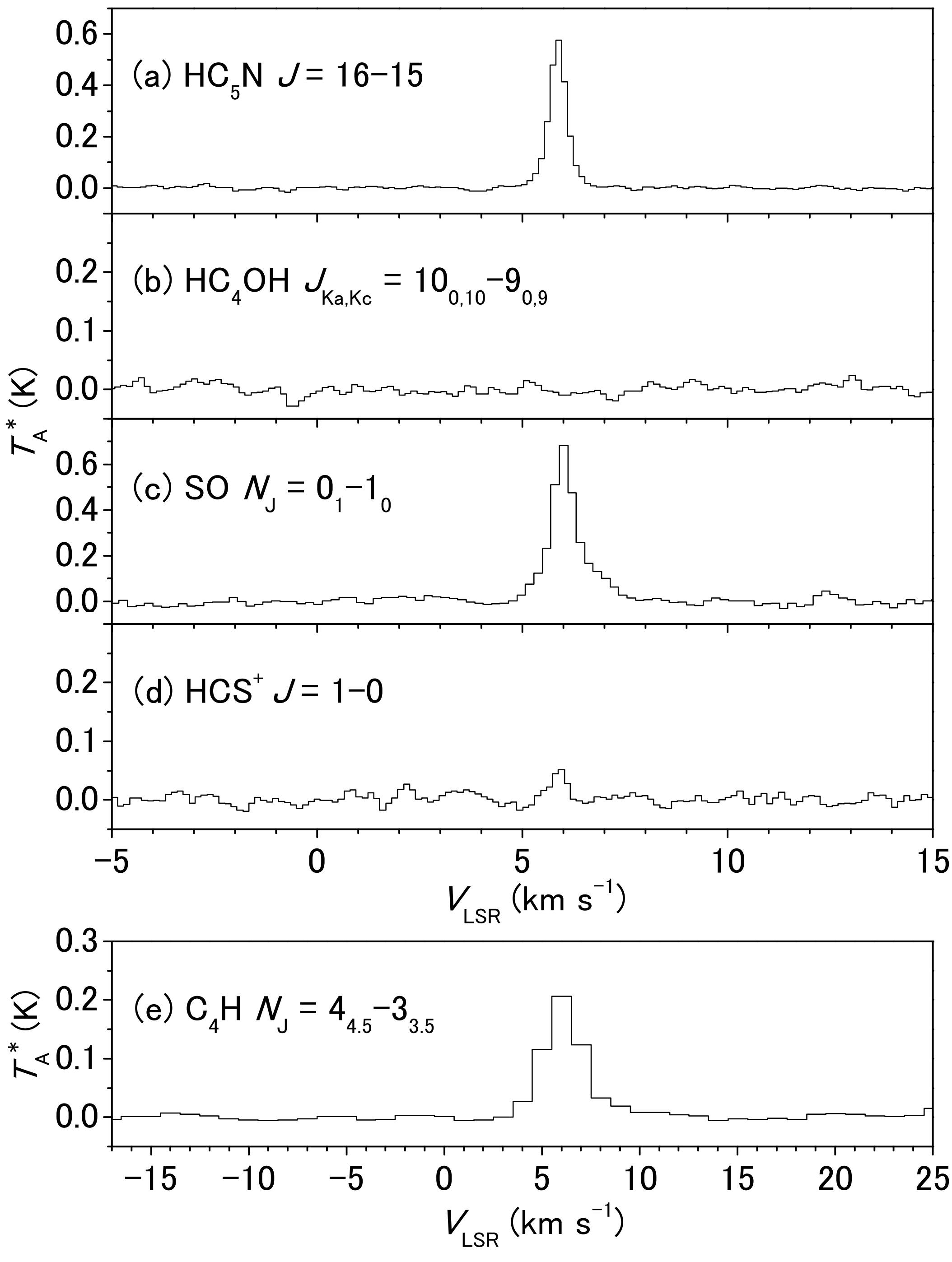

(a) HC5N J = 16–15
(b) HC4OH JKa,Kc = 10 0,10–9 0,9
(c) SO NJ = 01–10
(d) HCS+ J = 1–0
(e) C4H NJ = 4 4.5–3 3.5
T*A (K)
VLSR (km s−1)